\documentclass[12pt]{article}
\usepackage[utf8]{inputenc}
\usepackage{amssymb,amsmath,mathrsfs,enumerate}
\usepackage{graphicx,rotate,multicol}
\usepackage{float}
\usepackage{tocloft}
\usepackage{subfig}
\usepackage[margin=10pt,labelfont=bf]{caption}
\usepackage{cite}
\usepackage{soul}
\usepackage{slashed}
\usepackage{multirow}
\usepackage[colorlinks=true,
linkcolor=blue,
urlcolor=blue,
citecolor=teal]{hyperref}

\makeatletter
\let\Hy@linktoc\Hy@linktoc@page
\makeatother

\usepackage{color}
\definecolor{ourcolor}{rgb}{0.7, 0.25, 0.05}
\usepackage{tikz,braket}
\long\def\rpl#1!!#2!!{\textcolor{red}{#1} \textcolor{blue}{#2}}

\let\tilde=\widetilde
\let\hat=\widehat
\let\bar=\overline

\def \order(#1){{\mathcal O} \left(#1 \right)}

\evensidemargin=\oddsidemargin
\def\vev#1{\left\langle #1\right\rangle}

\allowdisplaybreaks 

\title{\color{black}{\bf Non-Standard Neutrino Interactions in a Modified $\nu$2HDM}}

\author {\bf Ujjal Kumar Dey,$^{a,}$\footnote{ujjal.dey1@gmail.com} 
	\hspace{4pt} Newton Nath,$^{b,c}$\footnote{newton@ihep.ac.cn} 
	\hspace{4pt} Soumya Sadhukhan$^{d}$\footnote{soumyas@prl.res.in} \\[10pt]
	\small\em $^a$Centre for Theoretical Studies, Indian Institute of Technology Kharagpur, Kharagpur 721302, India\\
	\small\em $^b$Institute of High Energy Physics, Chinese Academy of Sciences, Beijing 100049, China\\
	\small\em $^c$School of Physical Sciences, University of Chinese Academy of Sciences, Beijing 100049, China\\
	\small\em $^d$Physical Research Laboratory, Ahmedabad 380009, India
}

\date{}

\begin{document}

\maketitle

\begin{abstract}
In the traditional neutrinophilic two-Higgs doublet model ($\nu$2HDM), there is no non-standard neutrino interaction (NSI) as the interactions between the Standard Model fermions with neutrinos are negligibly small due to the tiny mixing of the two scalar doublets. In this work, we propose that if $\nu$2HDM is modified by considering the right-handed electron, $e_R$ is negatively charged under a global $U(1)$-symmetry then one can generate significant amount of NSI along with the tiny Dirac neutrino mass. 
Depending on different constraints from the LEP experiment, tree level lepton flavor violating processes, big-bang neucleosynthesis etc., we observe that this model significantly restricts the range of permissible NSI parameters, putting a strict upper bound on different NSIs. Furthermore, considering these model-dependent NSIs, we study their impact on the next-generation superbeam experiment, DUNE. We present a detailed discussion on the mass  hierarchy sensitivity  and the CP-violation discovery study considering the impact of both diagonal as well as off-diagonal NSIs. 
\end{abstract} 

\newpage

\hrule \hrule
\tableofcontents
\vskip 10pt
\hrule \hrule 

\section{Introduction}
\label{sec:intro}
The Standard Model (SM) of particle physics epitomizes our
knowledge of fundamental interactions of subatomic world with all 
its grandeur. Barring from a few minor disagreements the SM is 
hitherto untarnished by the direct observations from high energy 
colliders like LHC. The most elusive of the particles that 
constitute the SM is undoubtedly the neutrinos which can very well 
usher us towards the physics beyond the SM owing to its perplexing 
properties that they are massive (though tiny) and their different flavors are substantially mixed \cite{OHLSSON20161}.
These distinct attributes are exotic to the basic tenets of SM but are experimentally observed. 
The discovery of neutrino oscillation indubitably established the 
existence of mass for the neutrinos, whereas in the SM the 
neutrinos are supposed to be massless. The Super-Kamiokande 
experiment in Japan published their result to establish the 
phenomenon of neutrino oscillation in 1998 \cite{Fukuda:1998mi}. 
Later, various other experiments such as solar, reactor  and most 
recently long baseline experiments like, T2K \cite{Abe:2017uxa} and NO$ \nu $A 
\cite{Adamson:2017gxd} confirm the phenomenon of neutrino 
oscillation. In addition to the confirmation of the presence of  
neutrino mass, neutrino oscillation results bring forth some 
pertinent questions, e.g., 
(i) the neutrino mass hierarchy, i.e. whether neutrinos obey
normal hierarchy (NH, $m_3 > m_2 > m_1$) or inverted hierarchy 
(IH, $m_3 < m_1 \approx m_2$),  
(ii) the octant of $\theta_{23}$, i.e. $\theta_{23}$ lies in the
lower octant (LO, $\theta_{23} < 45^\circ$) or in the higher 
octant (HO, $\theta_{23} > 45^\circ$) and  
(iii) the determination of Dirac CP phase, $ \delta_{CP} $.
As the current experiments are striving to resolve the
degeneracies in hierarchy and octants and to assess the
sensitivity in CP violation, new theoretical inputs which can 
possibly help to address these issues are worth exploring. 
It is thus quite interesting to reconnoiter models which can
provide natural explanation to the tiny neutrino mass along 
with other issues mentioned above.
The enigma of neutrino mass lead the particle physicists to explore various theoretical models which can explain the neutrino mass as well as the observed neutrino mixing angles. Literature on the neutrino mass has been growing for quite some time, and today there are a number of phenomenological models addressing this issue (for a review see~\cite{King:2015aea, Cai:2017jrq}). 
%
Some of the new physics scenario could lead to corrections to the effective neutrino interactions through higher dimensional operator. Non-standard interactions (NSI) of neutrinos can be induced by the new physics beyond the SM (BSM). In the BSM scenarios NSI can arise when the heavier messenger fields are integrated out which can generate the dimension-6 \cite{Buchmuller:1985jz, Bergmann:1998ft, Bergmann:1999pk} and dimension-8 \cite{Berezhiani:2001rs, Davidson:2003ha} effective operators. For a detailed review and phenomenological consequences see Refs.~\cite{Ohlsson:2012kf, Miranda:2015dra} and the references therein. These were originally discussed even before the establishment of neutrino oscillation phenomena~\cite{Wolfenstein:1977ue, Valle:1987gv, Roulet:1991sm, Guzzo:1991hi}. 
Over the year many BSM scenarios have been studied where NSIs can be realized. Some of the popular models where this can be present are the $U(1)$ extended models with new $Z^{\prime}$ particle as the messenger, models with single or multiple charged heavy scalars, leptoquark, $R$-parity violating supersymmetry etc.~\cite{Farzan:2017xzy, Bilenky:1993bt, Barranco:2007tz, Antusch:2008tz, Malinsky:2008qn, Ohlsson:2009vk, Forero:2016ghr}.
Moreover, there are studies which considers the effect of NSI in collider experiments~\cite{Anchordoqui:2012qu, Franzosi:2015wha, Davidson:2011kr, Davidson:2011xz, Friedland:2011za, Choudhury:2018xsm}. Normally, the extensions of the SM that give rise to NSIs can have stringent constraints from the charged lepton flavor violating (LFV) processes. Therefore, it is imperative to maintain these LFV constraints while building any model from the NSI perspective.
We construct a variant of neutrinophilic two-Higgs doublet model ($\nu$2HDM) where we can have a significant NSI while maintaining various phenomenological constraints. In the standard $\nu$2HDM there is no NSI due to the absence of interaction between left-handed neutrinos with the other leptons and quarks via the extra scalars. We modify the standard $\nu$2HDM such that the second scalar doublet $\Phi_{2}$ couples only to the electron and neutrinos. This is achieved by assigning a negative charge of the $e_{R}$ under a global $U(1)$ symmetry. The charged Higgs present in the spectrum couples the left-handed electron and the charged leptons and thus can give rise to NSI by playing the role of the messenger. In this set-up we can have different NSI parameters. 
Apart from the presence of NSI the hierarchy in the fermion sector is somewhat less drastic in this model compared to the SM. This is because of the fact that the second Higgs doublet couples only to the electron and neutrinos and thus it is responsible for the mass of neutrinos and electron. The first Higgs doublet takes care of all the other SM fermions. Clearly, the hierarchy in the Yukawa couplings is minimized owing to the two vacuum expectation values (vev) of the two doublets. This betterment in the fermion hierarchy is one of the novel features of this model. Phenomenological constraints characteristic to any 2HDM will also be applicable in this scenario too. We consider the constraints on the model parameter space from lepton flavor violating processes (LFV), oblique parameters, $\mu_{g-2}$, big-bang neucleosynthesis (BBN) etc. 
Furthermore, we study the impact of model-dependent NSIs in case of  the long baseline (LBL) experiment like, DUNE. For DUNE, the mass hierarchy sensitivity considering model-independent bounds on NSI has been studied in~\cite{Liao:2016hsa,Coloma:2016gei,Deepthi:2016erc,Deepthi:2017gxg}. These studies show that the mass hierarchy sensitivity of LBL experiments get seriously compromised due to the presence of intrinsic $ \epsilon_{ee} \to - \epsilon_{ee} -2$, $ \delta_{CP} \to \pi - \delta_{CP} $ degeneracy. This degeneracy remains true irrespective of baseline and energy. On the other hand, if LBL experiment, like DUNE observes sensitivity then it will be able to rule out certain parameter space of model-independent NSIs. In the modified $\nu$2HDM model considered here, parameter space of different NSIs are constrained which helps to avert mass hierarchy degeneracy. We further examine the mass hierarchy sensitivity of DUNE considering model-based NSI parameters.
Later, we also address the issue of CP-violation (CPV) sensitivity for DUNE. We illustrate the role of new CP-phase on the measurement of $ \delta_{CP} $ considering both CP-conserving as well as CP violating values for the new phases. In our analysis, we assume NSI both in data as well as in theory. 
Some of the recent phenomenological studies, considering  model-independent  bounds of NSI, in the context of DUNE can be found in \cite{Masud:2015xva, deGouvea:2015ndi, Coloma:2015kiu, Liao:2016hsa, Soumya:2016enw, Blennow:2016etl, Forero:2016cmb, Huitu:2016bmb, Masud:2016bvp, Coloma:2016gei, Masud:2016gcl, Agarwalla:2016fkh, Liao:2016bgf, Fukasawa:2016gvm, Blennow:2016jkn, Liao:2016orc, Deepthi:2016erc, Ghosh:2017ged, Ghosh:2017lim, Deepthi:2017gxg} and the references therein.
We organize our paper as follows. In section~\ref{sec:nsi1}, we introduce the concept of non-standard neutrino interaction and their model-independent bounds. Section~\ref{sec:model} is devoted to a detailed description of the model that has been considered in this work. Traditional $\nu$2HDM model is discussed in 
the first part whereas the modification of the model which leads to NSI is discussed in 
the second part. Various phenomenological constraints coming from 
LEP data, LFV processes, BBN etc. are discussed in section
\ref{sec:phenoconstr}. We illustrate the effect of diagonal and 
off-diagonal NSI for DUNE in section~\ref{sec:nsi}. First, we 
present our result considering appearance channel probability ($ 
P_{\mu e} $). Later, we discuss the impact  of NSI on the 
determination of mass hierarchy sensitivity and the CPV 
sensitivity in case of DUNE. Finally, we summarize and conclude our noteworthy findings 
in section~\ref{sec:concl}. 

\section{General Description of NSI}
\label{sec:nsi1}
In this study, we consider the effect of neutral-current NSI in 
presence of matter which is describe by the dimension-6 four-
fermion operators of the form \cite{Wolfenstein:1977ue},
\begin{equation}
\label{eq:NSI}
\mathcal
{L}^{NC}_\text{NSI} = - 2\sqrt{2} G_F 
                      (\overline{\nu}_\alpha \gamma^{\rho} 
                       P_L \nu_\beta)
                      (\bar{f} \gamma_{\rho} P_C f)
                      \epsilon^{fC}_{\alpha\beta} 
                      + \text{h.c.}
\end{equation}
where $\epsilon^{f C}_{\alpha\beta}$ are NSI parameters, 
$\alpha, \beta = e, \mu, \tau$, $C = L,R$, $f = e, u, d$, and 
$G_{F}$ is the Fermi constant\footnote{Note that here we neglect 
NSI due to charges-current interactions which mainly affect 
neutrino production and detection \cite{Khan:2013hva, 
Ohlsson:2013nna, Girardi:2014kca, DiIura:2014csa, 
Agarwalla:2014bsa, Blennow:2015nxa}.}. Note that in general, 
the elements of $\epsilon_{\alpha\beta}$ are complex for 
$\alpha\neq \beta$ and real for $\alpha=\beta$ due to the 
hermitian nature of the interactions. For the matter NSI, 
$\epsilon_{\alpha\beta}$ are defined as,
\begin{equation}
 \epsilon_{\alpha\beta}  = \sum_{f,C}  \epsilon^{fC}_{\alpha\beta} \dfrac{N_f}{N_e} ~,
\end{equation}
where  $N_f$ is the number density of fermion $f$ and 
$\epsilon^{fC}_{\alpha\beta} = \epsilon^{f L}_{\alpha\beta} 
+ \epsilon^{f R}_{\alpha\beta}$. For the Earth matter, we can 
assume that the number densities of electrons, protons, and 
neutrons are equal (i.e. $ N_p \simeq N_n =N_e $), in such a case 
$ N_u \simeq N_d \simeq  3N_e $ and one can write,
\begin{equation}
\epsilon_{\alpha\beta} = \sqrt{\sum_C \left(
         (\epsilon^{e C}_{\alpha\beta})^{2} 
         + (3 \epsilon^{u C}_{\alpha\beta})^{2} 
         + (3\epsilon^{d C}_{\alpha\beta})^{2} 
         \right)}.
\label{nsiee1}
\end{equation}
The modified Hamiltonian in presence of propagation NSI, in the 
flavor basis, can be written as,
\begin{equation} 
\label{nsi_hamil}
H = \frac{1}{2E} \left[U
                  \text{diag}(0,\Delta m^2_{21},\Delta m^2_{31})
                  U^\dagger 
                  + {\rm diag}(A,0,0) 
                  + A \epsilon_{\alpha\beta}
                  \right]\,,
\end{equation}
where $U$ is the Pontecorvo-Maki-Nakagawa-Sakata (PMNS) mixing 
matrix~\cite{Patrignani:2016xqp}, $\Delta m^2_{ij}=m^2_i-m^2_j$ 
($ i<j = 1,2,3 $), $A\equiv 2\sqrt2 G_F N_e E$ represents the 
potential arising from standard interactions (SI) of neutrinos
in matter, and $ \epsilon_{\alpha\beta}\ $  can be written as
\begin{equation} 
\label{eq:potential}
\epsilon_{\alpha\beta}\ =  \left(\begin{array}{ccc}
 \epsilon_{ee} & \epsilon_{e\mu} & \epsilon_{e\tau} 
\\
\epsilon^{*}_{e\mu}  & \epsilon_{\mu\mu} & \epsilon_{\mu\tau}
\\
\epsilon^{*}_{e\tau} & \epsilon^{*}_{\mu\tau} & \epsilon_{\tau\tau}
\end{array}\right)\, ,
\end{equation}
where $ \epsilon_{\alpha\beta} = |\epsilon_{\alpha\beta}| e^{i\phi_{\alpha\beta}} $ for $ \alpha \neq\beta $.
The model-independent bounds \cite{Blennow:2014sja, 
Ohlsson:2012kf} on these parameters are 
\begin{align}
 & |\epsilon_{ee}|<4.2,~~|\epsilon_{e\mu}|<0.33,~~ |\epsilon_{e\tau}|<3.0\;, \nonumber  \\
 & |\epsilon_{\mu\mu}|<0.07,~~ |\epsilon_{\mu\tau}|<0.33,~~ \epsilon_{\tau\tau}|< 21 \;.
\end{align} 
Having introduced general descriptions of NSI and its model-independent bounds, in next section we describe our model and calculate the model-dependent bounds of different NSI parameters consistent with different experimental bounds.

\section{Model Description}
\label{sec:model} 
For completeness we first lay down the details of the traditional 
neutrinophilic two Higgs doublet model ($\nu$2HDM), for more 
details see e.g., \cite{Davidson:2009ha, Machado:2015sha} and 
references therein\footnote{For some other aspects of 2HDM see~\cite{Zarikas:1995qb, Lahanas:1998wf, Aliferis:2014ofa}}. In passing we note the reasons for the absence 
of NSI in the standard scenario. We then modify the vanilla 
$\nu$2HDM to get the NSI effects. 

\subsection{Standard $\nu$2HDM}
\label{sbsc:model1}
We start with an extension of the SM by adding an extra scalar doublet $\Phi_{2}$ having similar gauge quantum numbers as the SM scalar doublet $\Phi_{1}$. Three right handed neutrinos (RHN), $\nu_{Ri}$ are introduced and they will constitute the Dirac masses with the left handed neutrinos of the SM. There are mainly two ways to obtain the masses for the neutrinos with this particle spectrum. In one set-up~\cite{Ma:2000cc, Gabriel:2006ns} a $\mathbb{Z}_{2}$ symmetry is imposed under which the fields $\Phi_{2}$ and $\nu_{Ri}$ are odd and all the SM fields are even. In this set-up the Majorana mass terms for $\nu_{Ri}$ are not forbidden. In another set-up~\cite{Davidson:2009ha}, a global $U(1)$ symmetry is imposed and the fields $\Phi_{2}$ and $\nu_{Ri}$ have +1 charge under this and all the SM fields are neutral. In this scenario the neutrinos exclusively have Dirac masses. For the reasons that will be discussed in the later part we will follow the $U(1)$ case closely. In both of these cases the Yukawa interaction of the neutrinos is of the form $(-y_{\nu}^{ij}\bar{L}_{Li}\tilde{\Phi}_{2}\nu_{Rj})$ where $L_{L} = (\nu_{L},\ell_{L})^{T}$ is the left-handed lepton doublet and $\tilde{\Phi} = i\sigma_{2}\Phi^{\ast}$. The Yukawa structure of the other SM fermion is of the usual form involving $\Phi_{1}$. Note that when the $U(1)$ is unbroken, $\Phi_{2}$ has no vev and the neutrinos remain massless. 
The most general scalar potential for the exact $U(1)$ symmetric\footnote{If instead of the global $U(1)$ one imposes $\mathbb{Z}_{2}$ then another quartic term of the form $\lambda_{5}(\Phi_{1}^{\dagger}\Phi_{2})^{2}/2$ has to be added to $V(\Phi_1,\Phi_2)$.} case is given by
\begin{align}
\label{v2hdm}
V(\Phi_1,\Phi_2) &= m_{11}^2\Phi_1^\dagger\Phi_1 
                    + m_{22}^2\Phi_2^\dagger\Phi_2 
                     \nonumber  \\
                 & \quad + \dfrac{\lambda_1}{2}
                   (\Phi_1^\dagger \Phi_1)^2 
                   + \dfrac{\lambda_2}{2}
                   (\Phi_2^\dagger \Phi_2)^2 
                   + \lambda_3 \Phi_1^{\dagger} \Phi_1 
                                \Phi_2^{\dagger}\Phi_2
                   + \lambda_4 \Phi_1^{\dagger} \Phi_2 
                               \Phi_2^\dagger\Phi_1 \;.
\end{align}
The complex scalar $SU(2)$ doublets $\Phi_1$ and $\Phi_2$ carry hypercharge $Y=+1$. Since in the exact $U(1)$ symmetric case neutrinos are massless, to give masses to neutrinos this symmetry is to be broken. This is done by introducing a soft-breaking term of the form $(-m_{12}^{2}\Phi_{1}^{\dagger}\Phi_{2})$. Clearly, the smallness of $m_{12}^{2}$ technically natural in the 't-Hooft sense. The two scalar doublets can be presented as
\begin{equation}
\label{Higgs}
\Phi_a 	=
	\begin{pmatrix}
		\phi_a^+\\ (v_a+h_a+i\eta_a)/\sqrt{2} 
	\end{pmatrix},\qquad a=1,2.
\end{equation}
The vacuum expectation values (vev) of the two scalars can be denoted as $\vev{\Phi_1}=v_1$, $\vev{\Phi_2}=v_2$, and $v_{2}$ responsible for neutrino masses. Therefore, $v_{2}(\sim \text{eV})\ll v_{1}(\sim 246 \text{GeV})$ while $v^{2} = v_{1}^{2} + v_{2}^{2}$. In general the physical scalar fields are given by
\begin{subequations}
\begin{gather}
H^+ =\phi_1^+\sin\beta -\phi_2^+\cos\beta \simeq -\phi_2^+, \qquad  A =\eta_1\sin\beta - \eta_2\cos\beta \simeq -\eta_2,\\
h \;\;=-h_1\cos\alpha -h_2\sin\alpha \simeq -h_{1}, \qquad H =h_1\sin\alpha -h_2\cos\alpha \simeq -h_{2},
\end{gather}
\label{eq:phyScalar}
\end{subequations}
where $\alpha$ is the rotation angle for CP-even states and 
$\beta$ is of the CP-odd and charged scalar states. 
These rotations diagonalize the mass matrices with,
\begin{align}
\label{eq:alpha}
\tan 2\alpha &= \dfrac{2(-m_{12}^2 + \lambda_{34}\; v_1 v_2)}{m_{12}^2(v_2/v_1 - v_1/v_2) + \lambda_1 v_1^2 - \lambda_2 v_2^2} \;,\\ \label{eq:beta}
\tan\beta &= \dfrac{v_2}{v_1} \;,
\end{align}
where $\lambda_{34} \equiv \lambda_3 + \lambda_4$. From the 
requirement of tiny neutrino mass we have $v_2 \ll v_1$. 
Thus for small $m_{12}$, both $\alpha$ and $\beta$ will 
be very small. From Eqs.~\eqref{eq:phyScalar} it is evident that the SM-like 125 GeV Higgs comes from the doublet $\Phi_{1}$ whereas the BSM scalars are dominantly comprised of components of the doublet $\Phi_{2}$. The BSM scalars $H,A,H^\pm$ develop neutrinophilic interactions in the Yukawa sector. The Yukawa couplings of the new scalars in the limit $v_{2}\ll v_{1}$ are described as,
\begin{equation}
\label{yuk}
\mathcal{L}_{Y} \supset \frac{m_{\nu_i}}{v_2}H \bar{\nu_i}\nu_i 
               - i \frac{m_{\nu_i}}{v_2}
                   A \bar{\nu_i}\gamma_5\nu_i
               - \frac{\sqrt{2}m_{\nu_i}}{v_2} 
                  [U_{\ell i}^\ast H^+ \bar{\nu_i}P_L \ell
               + \mathrm{h.c.}],
\end{equation}
where $m_{\nu_i}$ are neutrino masses and $U_{\ell i}$ is the PMNS 
matrix. 
In this set-up due to the above-mentioned $U(1)$ charge assignments of the respective fields, left-handed neutrinos can not couple to right-handed charged leptons through the scalar doublet $\Phi_1$. The possibility of such coupling via the mixing  of the other doublet $\Phi_{2}$ is negligible since that coupling is proportional to $\sin \beta \approx v_2/v_1$ which is negligibly small. Since this type of coupling is needed to generate the NSI effect, in this set-up the NSI effect will be negligible.


\subsection{Modified $\nu$2HDM}
\label{sbsc:model2}
The standard $\nu$2HDM scenario discussed above can not give rise to non-standard interactions in the neutrino sector. Since two different Higgs doublets are responsible for providing mass to the neutrinos and other massive SM particles, there is no interaction of left-handed neutrinos with the SM massive leptons and quarks. Here we propose to give mass to the electron along with the neutrinos through the second scalar doublet $\Phi_2$. For that purpose, among the SM fields only the right-handed electron $e_R$ is endowed with the charge ($-1$) under the global $U(1)$. 
%
%
With this quantum number assignment the relevant terms for the neutrino and electron sector become
\begin{equation}
\label{nsi1}
\mathcal{L}^{m}_{\nu \text{2HDM}} \supset y_e \bar{L}_e \Phi_2 e_R + y_{\nu} \bar{L}_e \tilde{\Phi}_2 \nu_R + \text{h.c.},
\end{equation}
where $L_e$ is the SM electron doublet $(\nu_e \ \ e)^{T}_{L}$. 
Expanding the first term of the Lagrangian gives us terms 
involving charged Higgs ($H^{\pm}$) as,
\begin{equation}
\mathcal{L}^{\rm Yuk}_{H^{\pm}} \supset y_e \bar{\nu}_{eL} H^{+} e_R + \text{h.c.} 
\end{equation}
\begin{figure}[t]
 \begin{center}
   \includegraphics[width=5cm]{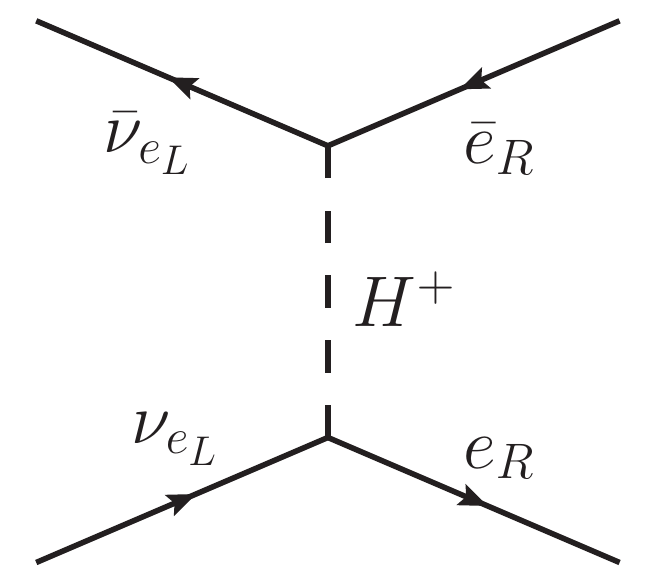}
 \end{center}
\caption{\footnotesize The Feynman diagram contributing to the NSI in the modified $\nu$2HDM.}
\label{fig:nsidiag}
\end{figure}
Presence of this interaction results in the $t$-channel process of 
Fig.~\ref{fig:nsidiag} which can give rise to non-standard 
interaction terms between SM neutrinos and electrons. 
The effective Lagrangian, after integrating out the heavy charged Higgs, looks like 
\begin{equation}
\mathcal{L}_{\rm eff} \supset  \frac{y_e^2}{4 m^2_{H^{\pm}}} (\bar{\nu}_{e L} \gamma^{\rho} \nu_{e L} ) \ (\bar{e}_R \gamma_{\rho} e_{R} )+ \text{h.c.}
\end{equation}
From this effective Lagrangian, comparing it with the defined form of the NSI Lagrangian, Eq.~\eqref{eq:NSI}, the NSI parameter $\epsilon_{ee}$ can be written as
\begin{equation}
\label{nsi}
\epsilon_{ee} =  \frac{1}{2\sqrt{2} G_F} \frac{y_e^2}{4 m^2_{H^{\pm}}}\;.
\end{equation}
Note that, in this model, only the right handed electron contributes in Eq.~\eqref{nsiee1} to provide $\epsilon_{ee} = \epsilon_{ee}^{eR}$.
With the same $U(1)$ quantum number assignment, the extra relevant terms for the lepton sector that can be added are:
\begin{equation}
\mathcal{L}^{m}_{\nu \text{2HDM}} \supset y_1 \bar{L}_{\mu} \Phi_2 e_R + y_2 \bar{L}_{\tau} \Phi_2 e_R + \text{h.c.},
\label{eq:lagnu2hdm}
\end{equation}
where $L_{\mu/ \tau}$ are the SM lepton doublets $(\nu_{\mu/ \tau} \ \ \mu/ \tau)^{T}_{L}$.
These terms will provide the following NSI parameters,
\begin{subequations}
\begin{align}
\epsilon_{e\mu} =  \frac{1}{2\sqrt{2} G_F} \frac{y_e y_1}{4 m^2_{H^{\pm}}} \;, \;\;\;\; 
\epsilon_{e \tau} =  \frac{1}{2\sqrt{2} G_F} \frac{y_e y_2}{4 m^2_{H^{\pm}}} \;,  \\
\epsilon_{\mu \tau} =  \frac{1}{2\sqrt{2} G_F} \frac{y_1 y_2}{4 m^2_{H^{\pm}}}  \;, \;\;\;\; 
\epsilon_{\mu \mu} =  \frac{1}{2\sqrt{2} G_F} \frac{y_1^2}{4 m^2_{H^{\pm}}} \;.
\end{align}
\end{subequations}
Note that here the Yukawa couplings $y_1, y_2$ can be complex.
The complex nature of these Yukawa results in generating phases 
for the NSI parameters. 
Before delving into the discussion on phenomenological constraints on this model and the NSI studies we mention a few features of this model. 
The first term of Eq.~\eqref{nsi1} appears due to modification in the $\nu$2HDM. The vacuum expectation value of second doublet $v_2$ gives mass to the electron here. Thus with an order one Yukawa coupling to get the electron mass $m_e= 0.51$~MeV we need $v_2 \sim$ MeV. With such a large $v_2 \sim$~MeV the neutrino Yukawas will be of the order of $10^{-6}$. This re-introduces hierarchy in the fermion Yukawa couplings i.e. the hierarchy of the Yukawas to accommodate mass of different fermions ranging from $\mathcal{O}({\rm eV})$ to $\mathcal{O}({\rm 100 \ GeV})$, with the same vev. 
In the SM if the neutrinos are to be given mass through the SM Higgs mechanism, a hierarchy $\mathcal{O}(10^{12})$ in the Yukawas is required to accommodate the top mass ($y_t \approx 1$) along with the Dirac neutrino ($y_{\nu} \approx 10^{-12}$) mass, using a vev of $246$~GeV. In the general $\nu$2HDM since the neutrinos are given mass through the second doublet, a hierarchy $\mathcal{O}(10^{6})$ is only required to arrange for masses from top quark to electron. 
In our modified $\nu$2HDM, we have a hierarchy $\mathcal{O}(10^{6})$ in Yukawa couplings to account for the mass of the electron and neutrinos using the same $v_2 \approx $~MeV. There remains another hierarchy $\mathcal{O}(10^3)$ to arrange for the masses of the rest of the fermions, i.e., from top quark to the muon, using the $\Phi_1$ vev $v_1 \approx 246$~GeV. Distributing the Yukawa hierarchies into two sectors, i.e. $\mathcal{O}(10^6)$ in $e-\nu$ sector and $\mathcal{O}(10^3)$ in $t-\mu$ sector, reduces the overall hierarchy of Yukawas in the model in comparison to the SM (where it is $\mathcal{O}(10^{12})$). 
However, the modified $\nu$2HDM has two different hierarchies, which is a somewhat less desirable feature than the requirement of only one hierarchy of order $10^6$ in the original $\nu$2HDM.
In this modified $\nu$2HDM, leptophilic BSM scalar couplings arise where $H$ ($A$) couples through the Lagrangian terms
\begin{equation}
\mathcal{L}_{LP} \supset \frac{y_e}{\sqrt{2}} H e_L e_R + i \frac{y_e}{\sqrt{2}} A e_L e_R + \text{h.c.}
\end{equation}
This coupling has stringent bounds from LEP data which we discuss in the next section.

Before ending this section we make a few remarks about the choice of $U(1)$-symmetric $\nu$2HDM over its $\mathbb{Z}_2$ symmetric counterpart.
Firstly, in the $\mathbb{Z}_2$ symmetric $\nu$2HDM CP-even new scalar $H$ is very light, $m_{H}\sim \mathcal{O}(v_{2}) \ll v$. For such a light scalar of mass up to MeV scale the constraint on the leptophilic coupling is very tight. If $\nu$2HDM with $\mathbb{Z}_2$ is allowed to have an electrophilic scalar coupling, the upper limit on the Yukawa $y_e$ will be of the order of $10^{-4}$. This strict bound on $y_{e}$ mainly comes from the $(g-2)_{e}$ as can be read off from Fig. 7 of~\cite{Knapen:2017xzo}. With a tiny Yukawa coupling which also determines the magnitude of the non-standard neutrino interaction, NSI effects of the modified model will be negligible. 
Secondly, the $\mathbb{Z}_2$ symmetric $\nu$2HDM has been severely constrained from the oblique parameters. The modification of the oblique $S$ parameter due to the presence of a very light neutral scalar, $m_H$, pushes it to a large negative value to rule out the model at $2 \sigma$ confidence level\footnote{It has been shown that the introduction of vector-like leptons relax this constraint~\cite{Sadhukhan:2018nsk}.}~\cite{Machado:2015sha}. In the $U(1)$ case, however, sufficiently heavy CP-even scalar is possible and also the mass degeneracy of CP-even and CP-odd scalars helps in satisfying $T$ parameter bounds easily.
%
%
%
\section{Phenomenological Constraints}
\label{sec:phenoconstr}
Typical to any 2HDM scenario the parameter space of the present set-up will also be subject to constraints from the data pertaining to heavy scalars, flavor violating issues etc. Note that in our model the second doublet couples only to the leptons and thus the constrains from the LEP and LFV is of more relevance. In this section we discuss constraints on the heavy scalars from the LEP searches, LFV decays, $(g-2)$ as well as BBN.

\subsection{Constraints from LEP}
\label{sbsc:LEP}

\subsubsection{Charged scalar mass}
The $Z$ boson decay width measurement at LEP points out that there is a tiny room for $Z$ decays to BSM particles. This suggests that the new scalars of the modified $\nu$2HDM should be heavier than half of $Z$ boson mass to kinematically forbid the decays. The search of a charged Higgs at LEP through the process $e^+ e^- \to Z \to H^+ H^-$ with $H^{\pm}$ further decaying to $\tau \nu$ puts a lower bound on the charged Higgs mass as $m_{H^{\pm}} > 80$~GeV~\cite{Abbiendi:2013hk}.
%

%
\subsubsection{Constraint from $e^+ e^- \to l^+ l^-$}
Measurement of $e^+ e^- \to e^+ e^-$ cross section at the LEP experiment can be expressed in terms of a limit on the scale of an effective four-electron interaction as $\lambda > 9.1$~TeV~\cite{LEP:2003aa}. In modified $\nu$2HDM, $e^+ e^- \to e^+ e^-$ process will take place through both the CP-even ($H$) and CP-odd ($A$) scalar mediators and the effective four lepton operator will look like 
\begin{equation}
\mathcal{L}_{\rm eff} \supset \frac{{y_e}^2}{8 m_H^2} (\bar{e}_L \gamma^{\rho} e_{L} )(\bar{e}_R \gamma_{\rho} e_{R} ) + 
\frac{{y_e}^2}{8 m_A^2} (\bar{e}_L \gamma^{\rho} e_{L} )(\bar{e}_R \gamma_{\rho} e_{R} ).
\end{equation} 
As in the case of a global $U(1)$-symmetric $\nu$2HDM, the degeneracy of CP-even and CP-odd scalar masses reduce the effective operator as,
\begin{equation}
\mathcal{L}_{\rm eff} \supset \frac{{y_e}^2}{4 m_H^2} (\bar{e}_L \gamma^{\rho} e_{L} )(\bar{e}_R \gamma_{\rho} e_{R} ), \nonumber
\end{equation} 
which being compared to the effective operator form with a scale $\Lambda$ provide the bound on $y_e$ as
%
$y_e^2  \le 8\pi m_H^2/ \Lambda^2.$
%
In our modified $\nu$2HDM with global $U(1)$ symmetry, the constraints coming from oblique parameters ($S, T$) allow mass splitting between the charged scalars and the neutral ones to be largest when $m_{H^{\pm}}$ is small ($\sim 100$~GeV)~\cite{Machado:2015sha}. For a light charged Higgs the neutral scalar masses can go up to several hundred GeVs. We take $m_H = m_A = 500$~GeV. If higher values of charged Higgs masses are taken then mass splitting between charged and neutral scalars decreases, which along with large $m_H^{\pm}$ constrains the Yukawa coupling tightly enough to allow very small values of NSI parameters. With $m_H = m_A = 500$~GeV and $\Lambda=9.1$~TeV the constraint translates to $y_e \le 0.28$ which further translates to a limit on the vev of $\Phi_2$ as
\begin{equation}
v_2  \ge  2.5 \ \text{MeV}.
\end{equation}
Using this bound we fix the $v_2$ value at $2.5~$MeV to get tightest limits on other Yukawas $y_1, y_2$ from LEP measurement in other processes like,  $e^+ e^- \to \mu^+(\tau^+) \mu^- (\tau^-)$. 
%
\subsubsection{Mono-photon constraint}
Another LEP constraint arises from the dark matter search in the mono-photon signal $e^+ e^- \to \text{DM}~\text{DM}~\gamma$, where $\gamma$ is either initial state radiation or it appears due to internal bremsstrahlung. In modified $\nu$2HDM similar mono-photon processes can occur, $e^+ e^- \to \nu_{e/\tau} \nu_{e/\tau} \gamma$ through the charged Higgs exchange which couples right-handed electron with left-handed neutrinos and vice-versa. The LEP DM search limit is thus rewritten as 
\begin{equation}
y_e^4 + 2 y_e^2 y_2^2 + y_2^4 \le \frac{16 m_{H_{\pm}}^4}{ \Lambda_{DM}^4},
\end{equation} 
with DM scale $\Lambda_{DM} \approx 320$~GeV~\cite{Fox:2011fx} for light DM particles.\footnote{A similar expression for $y_{1}$ can also be obtained. But we will see in the next section that a more stringent constraint on $y_{1}$ comes from lepton flavor violating decays.} 
%

\subsection{Lepton Flavor Violation Constraints}
\label{sbsc:lfvconstr}

\subsubsection{$\ell_\alpha \to\ell_\beta \gamma$}
For a generic process $\ell_\alpha \to\ell_\beta \gamma$, the scalar mediated branching ratio reads~\cite{Fukuyama:2008sz}
\begin{equation}
\mathrm{BR}(\ell_\alpha\to \ell_\beta\gamma) = \mathrm{BR}(\ell_\alpha\to e\bar{\nu}\nu)\frac{\alpha_\text{EM}}{192 \pi} |\langle m_{\alpha \beta}^2 \rangle|^2 \rho^2 \, .
\end{equation}
The strongest bound on this type of lepton flavor violating decay comes from the MEG experiment which gives the upper limit as $\mathrm{BR}(\mu\to e \gamma)<5.7 \times 10^{-13}$~\cite{Adam:2013mnn}, while bounds on the other LFV decay channels ($\tau \to e \gamma, \tau \to \mu \gamma$), obtained from the BaBar Collaboration, are weaker.  
Though the lepton flavor violating processes like $\mu \to e \gamma$ happen through the loop driven processes in this model, experimental constraints are strong enough to severely constrain this effect. The branching ratio for this decay which occurs through the charged scalar mediator can be written as~\cite{Bertuzzo:2015ada}
\begin{align}
\mathrm{BR}(\mu \to e \gamma) = \mathrm{BR}( \mu \to e\bar{\nu}\nu) \frac{\alpha_{\text{EM}}}{192 \pi} |\langle m_{\mu e}^2 \rangle|^2 \rho^2 \, ,
\end{align}
where $\rho = (G_F m_{H^\pm}^2 v_2^2)^{-1}$.
In terms of $\rho$ defined here, the $90\%$ confidence level bound read:
\begin{equation}
 \begin{array}{ll}
  \rho \lesssim 1.2 \, \mathrm{eV}^{-2} & ~~~~~[\mu \to e \gamma]\, , \\
 \end{array}
\end{equation}
The limit on $\rho$ translates into a limit where for $v_2 \lesssim 1$ eV 
one must have $m_H^{\pm}\gtrsim 250$ GeV. This is the tightest limit upto now on the $v_2$ and $m_{H^\pm}$ parameter space. Limits from similar other LFV decay modes are less constraining. 
With the sensitivity of the MEG expected to be improved further, the bound on $\rho$ is expected to be improved by about one order of magnitude. The limits imposed by the MEG bound on the $(m_{H^\pm}, v_2)$ plane are shown in Fig.~1 of \cite{Bertuzzo:2015ada}.
%

%
\subsubsection{$\tau (\mu) \to 3 e$}
The lepton flavour violating (LFV) decays play a crucial role in constraining the model parameters. From Eq.~(\ref{eq:lagnu2hdm}) it is evident that once $\Phi_{2}$ develops vev, there exist tree-level mixing among the would-be mass eigenstates $e,\mu$, and $\tau$. Therefore, LFV decays like, $\tau \to 3 e$ and $\mu \to 3 e$ are possible in this modified $\nu$2HDM through the neutral scalar ($H, A$) mediation at tree level and thus imply stringent constraints on the Yukawa couplings $y_1$ and $y_2$. Belle results~\cite{Hayasaka:2010np} of $\tau \to 3 e$ decay \textcolor{blue}{can be} normalized to $\tau \to \mu \nu_{\mu} \nu_{\tau}$ decay as,
\begin{equation}
\frac{\Gamma(\tau \to 3 e)}{\Gamma(\tau \to \mu \nu_{\mu} \nu_{\tau})} \le 1.58 \times 10^{-7}, \nonumber
\end{equation}
which implies the constraints on $\nu$2HDM parameter space as 
\begin{equation}
y_e y_2 \le \frac{(0.16 \ m_H)^2}{(1 \text{TeV})^2} \;.
\end{equation}
This is a tighter bound on $y_e$-$y_2$ plane compared to the LEP $e^+ e^- \to l^+ l^-$ and LEP mono-photon constraints. Similarly for $\mu \to 3 e$ decay there is even stronger experimental limit as $\text{BR}(\mu \to 3 e) \le 1 \times 10^{-12}$ will put bound on $y_e$-$y_1$ plane as
\begin{equation}
y_e y_1 \le \frac{(8.12 \times 10^{-3} \ m_H)^2}{(1 \text{TeV})^2} \;.
\end{equation}  
For moderate $y_e$ and allowed $m_H$ values this bound reduces the allowed Yukawa values to a tiny value ($y_1 \sim 10^{-6}$) which makes any NSI parameter involving $y_1$ like $\epsilon_{e\mu}, \epsilon_{\mu \tau}, \epsilon_{\mu \mu}$ insignificant. We have not considered this Yukawa coupling for our further analysis.
\begin{figure}[t]
\center
\includegraphics[scale=1.0]{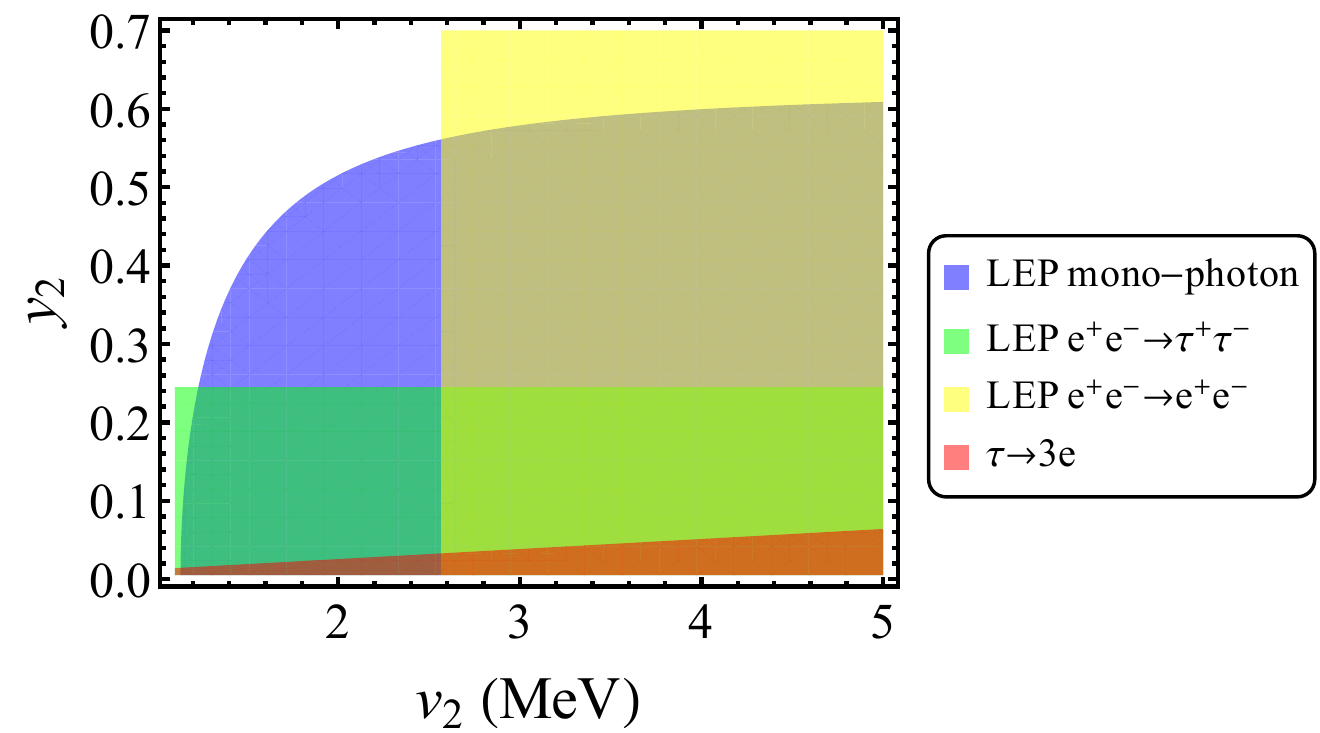}
\caption{\footnotesize Allowed region on the $y_2-v_2$ plane for different LEP constraints. 
Here $m_H, m_A = 500$~GeV, $m_{H^{\pm}} \sim 100$~GeV. 
This is the maximal mass difference allowed from $S, \ T, \ U$ and that maximal mass difference is crucial to generate large NSI values. }
\label{fig:allLEP}
\end{figure}
In Fig.~\ref{fig:allLEP} we show the $y_2-v_2$ parameter space allowed by various constraints from LEP and rare lepton decays.
Here the neutral scalar masses $m_H, m_A$ are kept at $500$~GeV where the charged Higgs mass is around $100$~GeV. 
As discussed earlier, maximal mass difference between neutral and charged scalars is required to enhance NSI parameter values. 
This is the amount of maximal mass splitting that is allowed from the $S, \ T, \ U$ parameters. 
\subsection{$g-2$ Constraints}
\label{sbsc:gmin2}
The charged scalars can contribute to muon and electron $g-2$~\cite{Dedes:2001nx} at one loop, but the contribution is negligible due to a suppression factor $m_l^4/m_{H^{\pm}}^2$ in the amplitude. Moreover, the loop contribution will depend on the coupling $y_{1}$ and since we have seen in the last section $y_{1} \sim 10^{-6}$, the contribution will be even more suppressed. Unlike a general 2HDM where the two loop contributions are dominant, charged lepton couplings to $H, A$ are suppressed here by a factor $\tan\beta$ in $\nu$2HDM, making $g-2$ constraints insignificant in this scenario, as shown in \cite{Grifols:1979yk}. 

%
\subsection{BBN Constraints}
\label{sbsc:bbn}
The new particles that are introduced in this model are the 
right-handed neutrinos which are very light, with mass of the 
scale of eV. These new degrees of freedom, ultra light neutrinos 
can be created in the early Universe to populate it through the 
process $\bar{l} l \to \nu_R \nu_R$ that occur through the 
exchange of the charged scalar $H^{\pm}$. From the big bang 
neucleosynthesis (BBN) there is a limit on new relativistic degrees of freedom which stands as 
$\Delta N_{\rm eff} \equiv N_{\rm eff} - 3.046 = 0.10^{+0.44}_{-0.43}$ 
at 95$\%$ confidence level with He + Planck TT + lowP + 
BAO data~\cite{Ade:2015xua}. From this one can obtain the decoupling temperatures of neutrinos which for right-handed neutrinos is $T_{d,\nu_{R}} \approx 200$~MeV and for left-handed ones $T_{d,\nu_{L}} \approx 3$~MeV~\cite{Olive:1999ij, Anchordoqui:2012qu, Zhang:2015wua}. These considerations put an  experimental upper limit on the charged Higgs mediated process $\bar{l} l \to \nu_R \nu_R$ as $(T_{d,\nu_{R}}/T_{d,\nu_{L}})^{3} 
\approx (\sigma_{L}/ \sigma_{R}) = 4(v_{2}m_{H^{+}}/(v_{1}m_{\nu_{i}}|U_{\l i}|))^{4}$~\cite{Davidson:2009ha}
which can be translated to a bound on the neutrino Yukawa coupling $y_{\nu_{i}}$ as
\begin{align}
y_{\nu_{i}} \le 0.05\times \left [\frac{m_{H^{\pm}}}{100~\text{GeV}}\right ] \left [\frac{1/\sqrt{2}}{|U_{ei}|}\right ].
\end{align}
In the usual $\nu$2HDM this constraint is used to put tight constraints on Yukawas. However, in the modified $\nu$2HDM this constraint is trivially satisfied due to the larger values of $v_2 \sim 0.1$~MeV; the preferred $y_{\nu_{i}}$ values are much  smaller.  

\vspace*{0.5cm}
In passing we also note that this kind of neutrinophilic models can be constrained by the limits on the effective four-neutrino interactions~\cite{Khlopov:1988f, Khlopov:1988s, Khlopov:1988t}. In $\nu$2HDM the effective 4$\nu$ interactions can occur through the neutral CP-even and CP-odd BSM scalar propagators. For a $\nu$2HDM with a global $U(1)$ symmetry the BSM neutral scalars will be heavier, with masses around $100$~GeV, which suppresses the effective interaction rate. Therefore, even with $H(A) \nu \nu$ Yukawa couplings $\sim$1, the effective four-neutrino vertex strength remains sufficiently low to stay below the experimental limits on effective four-neutrino interaction.

%
\section{Study of NSI}
\label{sec:nsi}
In this section, we first present a discussion on the constraint range of NSI parameters in this model. Furthermore, we proceed with a detailed study of these bounds considering the data from DUNE. Our initial discussion is on the impact of model-dependent NSI considering appearance channel probability for both neutrinos and antineutrinos. Moreover, we also perform substantially detailed study at the $\chi^{2}$ level where we illustrate the impact of NSI on mass hierarchy sensitivity and CP violation discovery study.

\subsection{NSI in the modified $\nu$2HDM}
\begin{figure}[tb]
\centering
 \subfloat[]{
   \includegraphics[width=7.1cm]{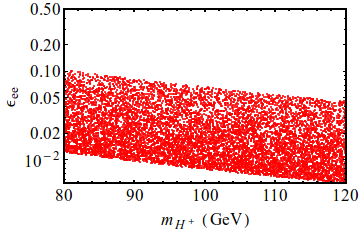}}
   \hspace{0.2cm}
 \subfloat[]{  
   \includegraphics[width=7.1cm]{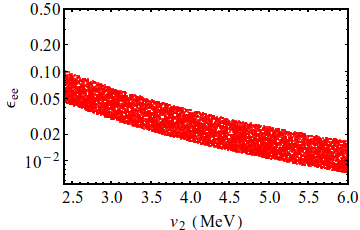}}
\caption{\footnotesize Allowed values of NSI parameter $\epsilon_{ee}$ for a random scan of parameters and its distribution with (a) charged Higgs mass, $m_{H^{\pm}}$, and (b) the vacuum expectation value of $\Phi_2$, $v_2$ (right).}
\label{fig:three}
\end{figure}

Here we fix different benchmark values of $v_2$ and vary $m_{H^{\pm}}$ in the range of $80-120$~GeV. The LEP bound sets the lower limit of charged Higgs mass at 80 GeV. To satisfy the LEP $e^+ e^- \to l^+ l^-$ constraint we need to have heavier electrically neutral BSM scalars as that can keep the cross section low. On the other hand, the heavier charged Higgs will suppress the values of non-standard neutrino interaction parameters, as can be seen from Eq.~\eqref{nsi}. Therefore, it is optimal to have a maximal splitting between charged and neutral BSM scalars and the splitting is controlled by oblique parameter constraints. Maximal splitting is possible for smaller mass of charged scalar upto $m_H^{\pm}=120$~GeV, when the neutral scalar masses are at $500$~GeV. So, the upper limit of $m_{H^{\pm}}$ is chosen at $120$~GeV. From the LEP upper limit on $e^+ e^- \to e^+ e^-$ process, a lower bound on $v_2$ value can be obtained as $v_2 \le 2.5$~MeV. For three different values of $v_2 \ge 2.5$~MeV we choose three benchmark points and calculate $\epsilon_{ee}$ for those values. The allowed values of $\epsilon_{ee}$ for various model parameters $m_{H^{+}}$ and $v_2$ are shown in Fig.~\ref{fig:three}. From this, we take some representative benchmark points for $\epsilon_{ee}$ along with $v_2$ and $m_{H^{+}}$ which are given in Table~\ref{nsiee}.
Constraints from lepton flavor violating decay $\mu \to  3 e$ forces the Yukawa coupling $y_1$ to be small, i.e. of the order of $10^{-6}$ so that all the $\epsilon$ values involving $\mu$ i.e. $\epsilon_{e\mu}, \epsilon_{\mu \mu}, \epsilon_{\mu \tau} $ will be negligible to have any significant effect on the observables. So we do not consider those NSI parameters for our analysis. The only NSI parameters that we explore are $\epsilon_{e\tau}$ and $\epsilon_{\tau \tau}$. In general, the model-independent constraint on NSI parameter $ \epsilon_{\tau \tau}$ is too loose to put any significant constraint on model parameters.
For the case of $y_2$, the tightest bound comes from LFV decay $\tau \to 3 e$ which restricts $y_2$ to be below $0.035$ for $v_2= 2.5$~MeV and this upper limit increases with $v_2$. We have chosen different values of Yukawa coupling $y_2$ within this limit. For these different $y_2$ values, $\epsilon_{e \tau}$ numbers at three benchmark points are given in Table~\ref{tab:nsiparam}. 
\begin{table}
\begin{center}
\begin{tabular}{ |c||c|c|c| } 
 \hline
 Parameters & Benchmark Point-I & Benchmark Point-II & Benchmark Point-III\\ \hline \hline 
 $v_2$ & 2.5 MeV  & 3 MeV & 5 MeV \\ 
 $m_{H^{\pm}}$ & 80-100 GeV &  80-100 GeV & 80-100 GeV\\ 
 $\epsilon_{ee}$ & 0.061-0.095 & 0.042-0.066 & 0.015-0.024 \\
 \hline
\end{tabular}
\vspace{0.2cm}
\caption{\footnotesize {\bf Benchmark Points and $\epsilon_{ee}$}: NSI parameter $\epsilon_{ee}$ for different benchmark values of $v_2$ where the charged Higgs mass is varied over a range 
from 80 GeV to 100 GeV (keeping in mind mass splitting allowed from oblique parameter considerations and mass splitting required for satisfying LEP limit) for each $v_2$ values.}
\label{nsiee}
\end{center}
\end{table}
\begin{table}[]
\centering
\begin{tabular}{|c|c|c|c|}
\hline 
\multicolumn{4}{|c|}{BP-I ($v_{2} = 2.5$ MeV)}                                                                                                                       \\ \hline
 $y_{2}$      & 0.01                              & 0.02                                                 & 0.035         \\ \hline
$\epsilon_{e \tau}$   & 0.0021 - 0.0033                           & 0.0043 - 0.0067                                                 & 0.0075 - 0.0117     \\ \hline
\multicolumn{4}{|c|}{BP-II ($v_{2} = 3.0$ MeV)}                                                                                                                      \\ \hline \hline
 $y_{2}$      & 0.01                              & 0.02                                                  & 0.04          \\ \hline
$\epsilon_{e \tau}$   & 0.0018 - 0.0028                           & 0.0036 - 0.0056                                                  & 0.0071 - 0.011      \\ \hline
\multicolumn{4}{|c|}{BP-III ($v_{2} = 5.0$ MeV)}                                                                                                                     \\ \hline \hline
$y_{2}$      &  0.02                              & 0.04                                                 & 0.065           \\ \hline
$\epsilon_{e \tau}$   & 0.0021 - 0.0033                           & 0.0043 - 0.0067                                               & 0.007 - 0.011    \\ \hline
\end{tabular}
\caption{\footnotesize NSI parameter $\epsilon_{e \tau}$ for three benchmark points (BP-I,II,III) for different values of $y_2$.}
\label{tab:nsiparam}
\end{table}
%

\subsection{Probability level descriptions}
We now study the appearance channel ($ P_{\mu e} $) probability in case of DUNE considering model-dependent NSI parameters, 
$ \epsilon_{ee} $ and $ \epsilon_{e \tau} $. 
The relevant analytical expression for the appearance channel probability in case of normal hierarchy (NH) considering $s_{13}$, $r = \Delta m^2_{21}/\Delta m^2_{31}$ and $\epsilon_{\alpha \beta}$ as a small parameters, except $ \alpha, \beta = e$, can be written as \cite{Liao:2016hsa},
\begin{eqnarray}
P_{\mu e} & = & x^2 f^2 + 2xyfg \cos(\Delta + \delta_{CP}) + y^2 g^2
\nonumber\\
&+& 4\hat A \epsilon_{e\tau} s_{23} c_{23}
\left\{ xf [f \cos(\phi_{e\tau}+\delta)  
- g \cos(\Delta+\delta+\phi_{e\tau})] \right.
 \left. -yg [g \cos\phi_{e\tau} - f \cos(\Delta-\phi_{e\tau})]\right\}
\nonumber\\
&+& 4 \hat A^2 (g^2 + f^2) c_{23}^2 s_{23}^2|\epsilon_{e\tau}|^2 - 8 \hat A^2 fg s_{23} c_{23}c_{23}\epsilon_{e\tau}^2 \cos\Delta 
\nonumber\\
&+& {\cal O}(s_{13}^2 \epsilon, s_{13}\epsilon^2, \epsilon^3)\,,
\label{eq:prob}
\end{eqnarray}
where
\begin{eqnarray}
&& x = 2 s_{13} s_{23},~
y = 2r s_{12} c_{12} c_{23},~
 (s_{ij} = \sin\theta_{ij},c_{ij}=\cos\theta_{ij}, ij = 12, 13, 23 ) \;,
\nonumber\\
\Delta &=& \frac{\Delta m^2_{31} L}{4E},\ 
\hat A = \frac{A}{\Delta m^2_{31}},\
f,\, \bar{f} = \frac{\sin[\Delta(1\mp\hat A(1+\epsilon_{ee}))]}{(1\mp\hat A(1+\epsilon_{ee}))}\,,\ 
g = \frac{\sin[\hat A(1+\epsilon_{ee}) \Delta]}{\hat A(1+\epsilon_{ee})}
\label{eq:define}
\end{eqnarray}
Similar expression for the inverted hierarchy (IH) can be calculated by replacing 
$\Delta m^2_{31} \to - \Delta m^2_{31}$ (i.e.  $ \Rightarrow \Delta \to - \Delta$,  $\hat A \to - \hat A$ 
(i.e. $f \to - \bar{f}$ and $g \to -g$), $ y \to -y $ ). 
Also the expressions for  antineutrino probability ($ P_{\overline{\mu} \overline{e}} $) can be obtained by replacing  $\hat A \to - \hat A$ ( $ \Rightarrow f \to \bar{f}$), $\delta_{CP} \to - \delta_{CP}$. 

%
DUNE is proposed as a next generation superbeam experiment at Fermilab, USA~\cite{Adams:2013qkq, Acciarri:2016crz, Strait:2016mof, Acciarri:2016ooe, Acciarri:2015uup}. It will use existing NuMI (Neutrinos at the Main Injector) beamline design at Fermilab as a source of neutrinos. Its far detector will be placed at Sanford Underground Research Facility (SURF) in Lead, South Dakota, at a distance of 1300 km from the source. DUNE collaboration has planned to use LArTPC (liquid argon time-projection chamber) detector with the volume of 10 kton and 40 kton corresponding to the first phase and final phase respectively.
In our simulation, we consider 10 kton detector volume. We also consider the flux corresponding to 1.2 MW beam power with 120 GeV protons which give $1\times 10^{21} $ protons on target (POT) per year. In the analysis, we use method of pulls~\cite{GonzalezGarcia:2004wg, Fogli:2002pt} as outlined in~\cite{Gandhi:2007td} to account for systematic errors. For the numerical simulation, we use the \texttt{GLoBES} package \cite{Huber:2004ka, Huber:2007ji} along with the required auxiliary files~\cite{messier_xsec, Paschos:2001np}.
The remaining details of the DUNE and numerical specifications considered here are the same as~\cite{Nath:2015kjg}. We have also added $5\%$ prior on sin$^{2}2\theta_{13}$.

\begin{figure}[!h]
\begin{center}
 \begin{tabular}{lr}
\hspace{-1.8cm}
\includegraphics[height=6.5cm,width=10cm]{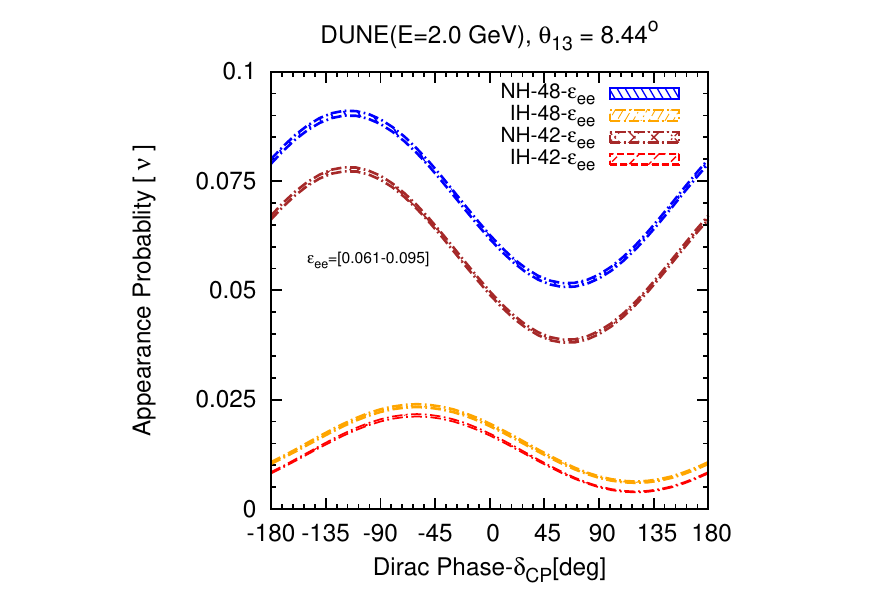}
  \hspace{-3cm}
\includegraphics[height=6.5cm,width=10cm]{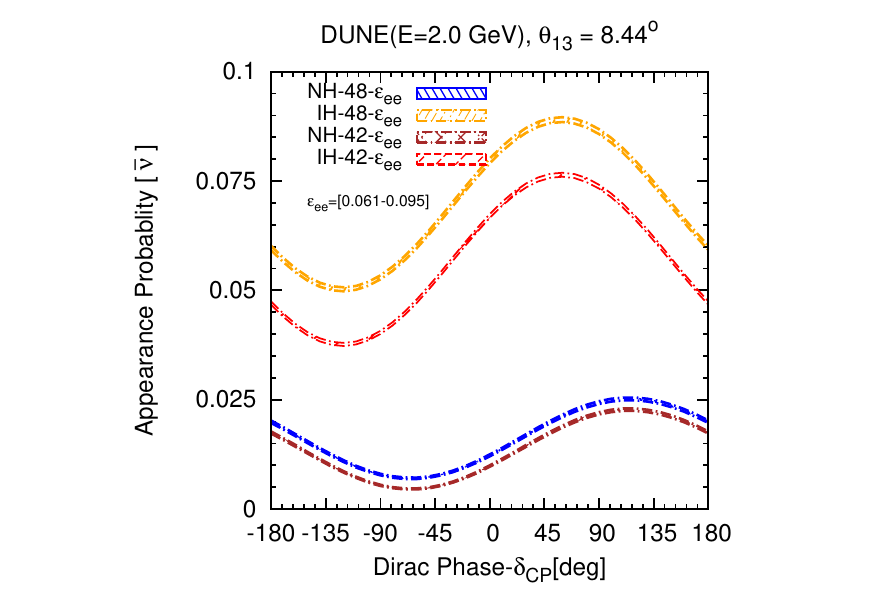} \\
  \hspace{-1.8cm}
\includegraphics[height=6.5cm,width=10cm]{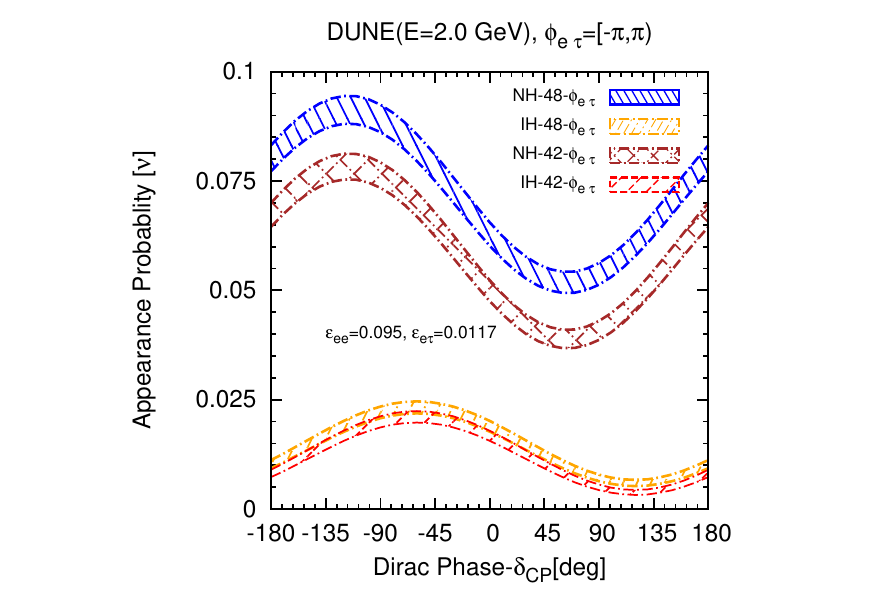}
  \hspace{-3cm}
\includegraphics[height=6.5cm,width=10cm]{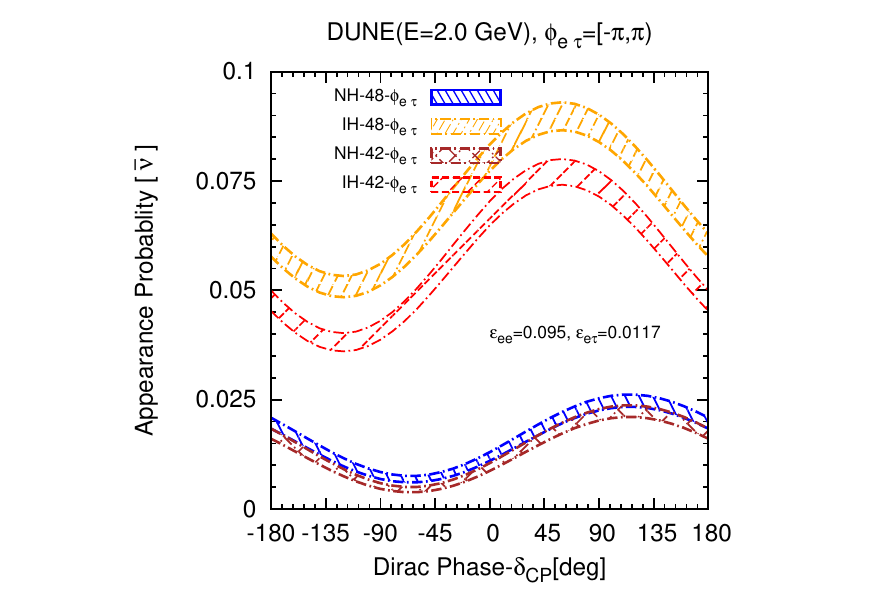} \\
 \end{tabular}
 \end{center}
\vspace{-4ex}        
\caption{\footnotesize Appearance channel probability vs. $\delta_{CP}$ for DUNE considering neutrinos (antineutrinos) in left (right) panel. Here top (bottom) row represents the oscillation probability in presence of diagonal (off-diagonal) NSI parameter $ \epsilon_{e e} $ ($ \epsilon_{e \tau} $).}
\label{fig:prob_ee}
\end{figure}
In Fig.~\ref{fig:prob_ee}, we discuss appearance channel probability ($ P_{\mu e} $) considering diagonal NSI parameter $ \epsilon_{ee} $ in the top row whereas in the bottom row, we show our results considering off-diagonal NSI parameter $\epsilon_{e \tau} $ considering 2 GeV neutrino beam energy. We present our results considering neutrinos (antineutrinos) in left (right) panel. Depending on true hierarchy and true octant, we have four combinations of hierarchy - octant namely, NH-HO, NH-LO, IH-HO and IH-LO. We discuss our results for all the four possibilities. The tiny bands in top row are over the range of model-dependent NSI parameter $ \epsilon_{ee} $ as given by the first  column of Table~\ref{nsiee}~\footnote{Note that for illustrative purpose, we describe our results considering one set of solution.}. 
In bottom row, we describe the role of off-diagonal NSI parameter where different bands are over the range of new CP-phase $ \phi_{e \tau} $ (see figure legends for details). Note that in the bottom row, to have better understanding of new CP phase, we fix the value of NSI parameters and the representative value of $ \epsilon_{e \tau} $ is taken from Table~\ref{tab:nsiparam}. 
A general probability-level discussion, considering similar set-up, in the absence of NSI on various degenerate solutions are noticed in Ref.~\cite{Nath:2015kjg} whereas some noteworthy issues considering model-independent NSI parameters are presented in Ref.~\cite{Deepthi:2016erc} considering DUNE.
We concentrate here on the impact of model-dependent NSI parameters. In the literature(see \cite{Liao:2016hsa, Coloma:2016gei, Deepthi:2016erc, Deepthi:2017gxg}), it was pointed out that appearance channel probability suffers from degeneracy due to the transformation of the form $({\rm NH}, \epsilon_{ee}) \to ({\rm IH}, - \epsilon_{ee} -2$), {\rm and } $({\rm NH}, \delta_{CP}) \to ({\rm IH}, \pi - \delta_{CP}) $ in presence of model-independent NSI parameters. Here we notice considering our model-dependent constrained parameter space of $ \epsilon_{ee} $ that DUNE has no hierarchy degeneracy as shown in the first plot of top row in Fig.~\ref{fig:prob_ee}. This can be understood by comparing NH bands (blue, brown) with IH bands (yellow, red), as the former have no intersection with latter. The underlying reason for the absence of this degeneracy can be traced back to the fact that the transformation required for this degeneracy, $({\rm NH}, \epsilon_{ee}) \to ({\rm IH}, - \epsilon_{ee} -2)$, can not be realized in our model as $\epsilon_{ee}$ is always positive in this model, as can be seen from Eq.~(\ref{nsi}). From the first plot of bottom row, we also observe that $ P_{\mu e} $ has no hierarchy degeneracy even in the presence of off-diagonal NSI parameter, $\epsilon_{e\tau}$. Note that with the inclusion of $\epsilon_{e\tau}$, width of different bands become boarder, this is due to unconstrained range of new CP-phase $\phi_{e\tau}$ which affects the measurement of Dirac CP-phase. Similar results are also observed for antineutrinos as shown by right panel and conclusion made for neutrinos remain the same for antineutrinos. Moreover, to understand the impact of NSI on the determination of hierarchy and CP violation sensitivity, we extend our study considering our model-dependent NSI parameters at the $ \chi^{2} $ level. In next section, we illustrate our details.

\subsection{Sensitivity study}
\begin{figure}[t]
\begin{center}
 \begin{tabular}{lr}
\hspace{-1.8cm}
\includegraphics[height=6.5cm,width=10cm]{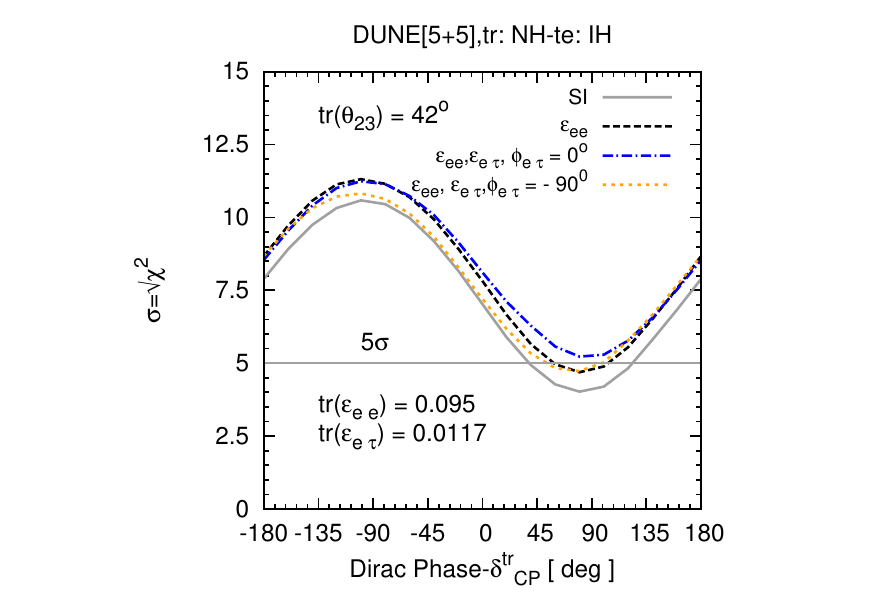}
\hspace{-3cm}
\includegraphics[height=6.5cm,width=10cm]{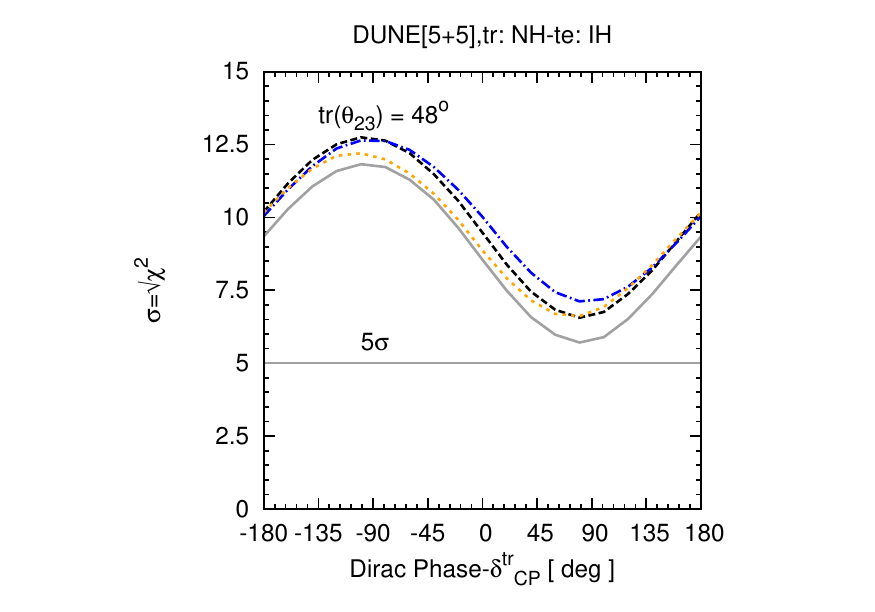}\\
\hspace{-1.8cm}
\includegraphics[height=6.5cm,width=10cm]{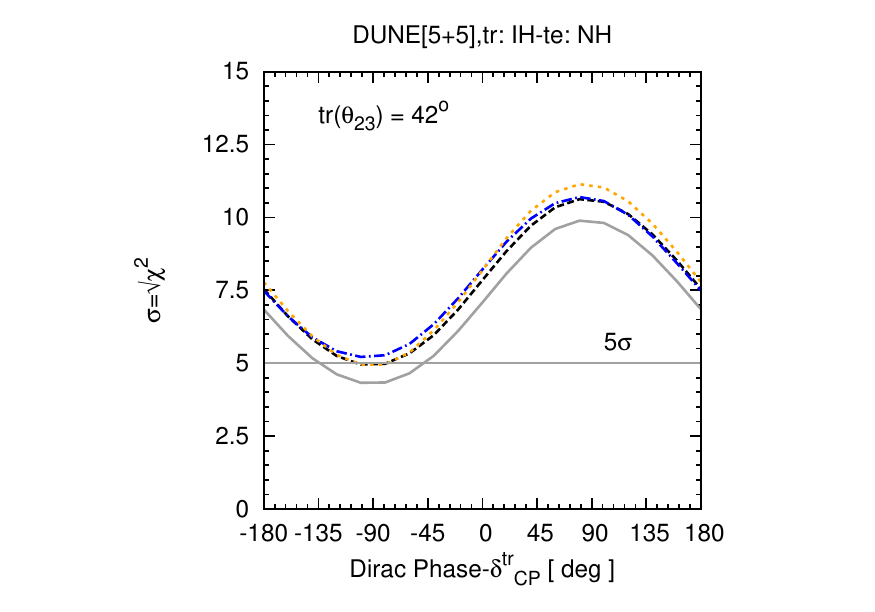}
\hspace{-3cm}
\includegraphics[height=6.5cm,width=10cm]{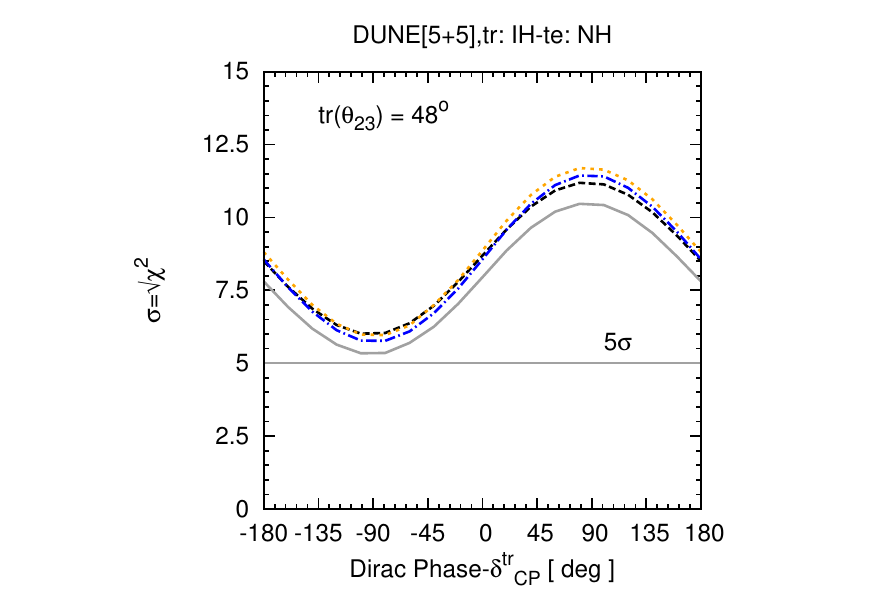}
 \end{tabular}
 \end{center}
\vspace{-4ex}        
\caption{\footnotesize Hierarchy $ \chi^{2} $ for DUNE in presence of NSI. Here top (bottom) row is for true NH (IH) and left (right) column is for LO (HO).}
\label{fig:hierarchy}
\end{figure}

We now discuss the mass hierarchy sensitivity as well as the CP-violation sensitivity of DUNE with model-dependent NSI parameters. For the comparison, we also describe our results considering the standard interactions (SI). Throughout the study, we consider [5+5] years run time for DUNE\footnote{Note that [5+5] means we divide total exposures with 5 years of neutrino run and another 5 years of antineutrino run.}. 
In Fig.~\ref{fig:hierarchy}, we describe our results for the mass hierarchy sensitivity considering true $\theta_{23}$ as $ 42^\circ$ for lower octant (LO) and $ 48^\circ $ for higher octant (HO) in the first and second panel respectively. Whereas in top (bottom) row, we present our results considering true hierarchy as NH (IH) and marginalized over test hierarchy.
We take true values of other neutrino parameters as, $\sin^{2} \theta_{12} = 0.321 $, $\sin^{2} 2\theta_{13} = 0.085$, $ \Delta m^{2}_{31}  = 2.50 \times 10^{-3}$ eV$ ^{2} $ and $ \Delta m^{2}_{21}  = 7.56 \times 10^{-5}$ eV$ ^{2} $ which are compatible with the current global-fit data~\cite{deSalas:2017kay, Capozzi:2016rtj, Gonzalez-Garcia:2015qrr}.
In test hierarchy, we marginalize over their $ 3\sigma $ ranges\footnote{Note that we do not marginalize over $\theta_{12}  $.}. For the NSI parameters, we marginalize over test $ \epsilon_{ee} $ considering the range as given in BP-I of Table~\ref{nsiee} whereas we keep fixed value for the off-diagonal NSI parameter as given in Table~\ref{tab:nsiparam} (see figure legends for details). This benchmark point is chosen since the respective $v_2$ value satisfies all other constraints and is well within the allowed  parameter region.  
Note that the four different curves correspond to SI (gray solid), diagonal NSI parameter $ \epsilon_{ee} $ (black long-dotted), off-diagonal NSI parameter $ \epsilon_{e \tau} $ with $\phi_{e\tau}=0^\circ$ (blue dash-dotted) and  $\phi_{e\tau}= - 90^\circ$ (yellow dotted) respectively. We quantify our mass hierarchy sensitivity as below,
\begin{equation}\label{eq:mass_hier}
\chi^{2}_{\rm NH-IH} = {\rm min} \sum_i \dfrac{ \left[  N_i({\rm NH}^{\rm tr}, \epsilon^{\rm tr}, \phi^{\rm tr})  - N_i({\rm IH}^{\rm te} , \epsilon^{\rm te}, \phi^{\rm tr}) \right] ^{2}}{ \sigma [N_i(  {\rm NH}^{\rm tr}, \epsilon^{\rm tr}, \phi^{\rm tr} )]^{2}} \; ,
\end{equation}
where $ N_i $ represents the number of events for the $i^{th}$ oscillation parameters. Also, $ \epsilon \equiv \epsilon_{\alpha \beta} $ are marginalized for $ \alpha = \beta = e$ and kept fixed in both true and test for $ \alpha \neq \beta $ whereas $ \phi \equiv \phi_{\alpha \beta} $ are considered fixed in both true and test to calculate $N$. 
We discuss our results considering a benchmark of $ 5\sigma $ confidence level (C.L.) as shown by the horizontal line.
From the gray curve, we observe that DUNE can reach $ 5\sigma $ hierarchy sensitivity with the 10 kton detector mass, independent of the true values of $\delta_{CP} $ and irrespective of the nature of true hierarchy NH or IH, in case of HO (right panel) for SI. Whereas for LO (left panel), we notice it achieves $ 5\sigma $ hierarchy sensitivity for all the true values of $ \delta_{CP} $ except the regions around true $ \delta_{CP} = + 90^\circ$ for neutrinos.
In the presence of $ \epsilon_{ee} $ (see black long-dotted curve), we find that DUNE achieves greater than or almost equivalent to $ 5\sigma $ sensitivity for all the mentioned cases. Considering $ \epsilon_{e \tau} $, we describe our results for both CP conserving (i.e., when $ \phi_{e \tau}  = 0^\circ $) as well as CP violating (i.e., when $ \phi_{e \tau}  = -90^\circ $) values of new CP-phase. 
In the case of HO, we notice that even in the presence of off-diagonal NSI parameters DUNE can achieve $5\sigma $ hierarchy sensitivity for all the true values of $\delta_{CP} $. Whereas in case of LO, we find that DUNE attains $ 5\sigma $ hierarchy sensitivity irrespective of the true values of $ \delta_{CP} $ for all the cases except the regions around true $ \delta_{CP} = + 90^\circ$. Similar results hold for the case of true IH as shown by bottom row, except that for IH, the  minimum is near true $ \delta_{CP} = - 90^\circ$. 
%

\begin{figure}[t]
\begin{center}
 \begin{tabular}{lr}
\hspace{-1.8cm}
\includegraphics[height=6.5cm,width=10cm]{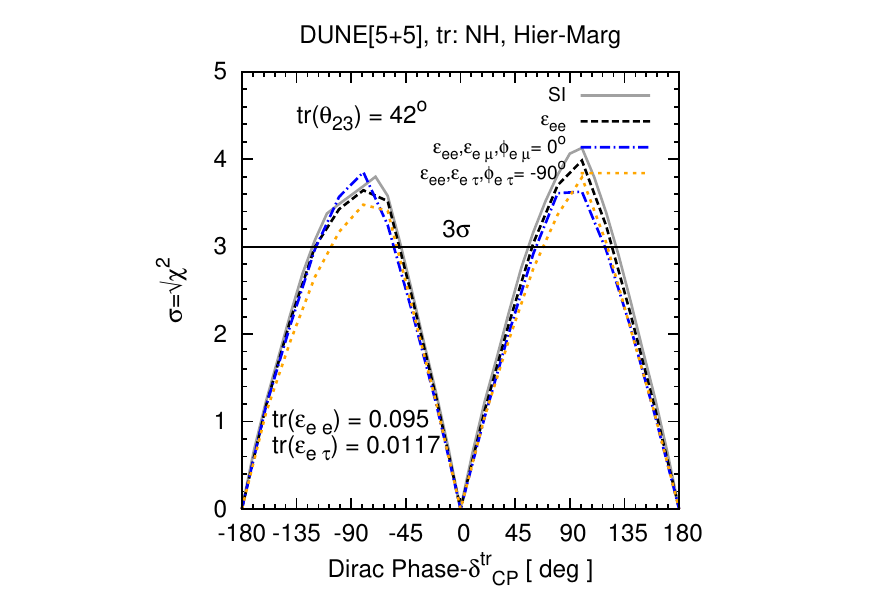}
\hspace{-3cm}
\includegraphics[height=6.5cm,width=10cm]{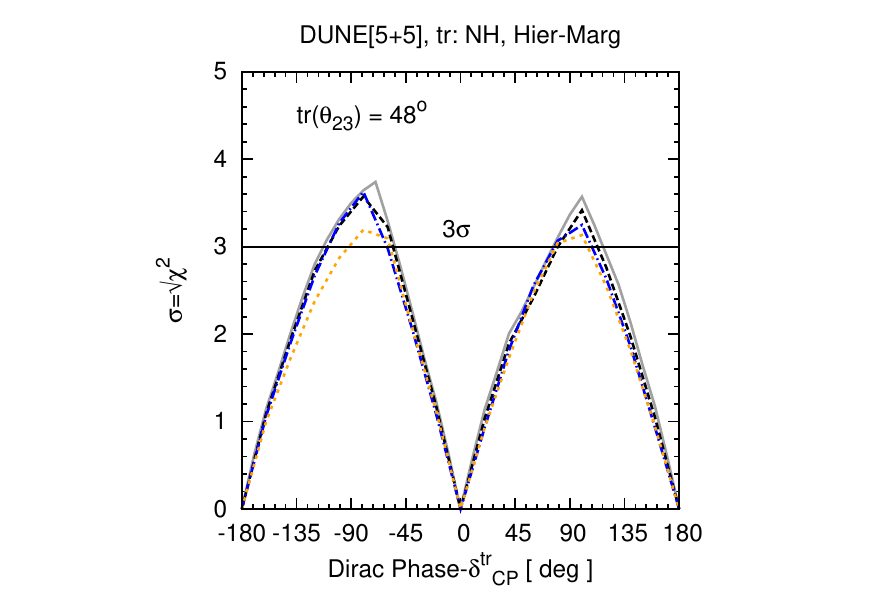} \\
\hspace{-1.8cm}
\includegraphics[height=6.5cm,width=10cm]{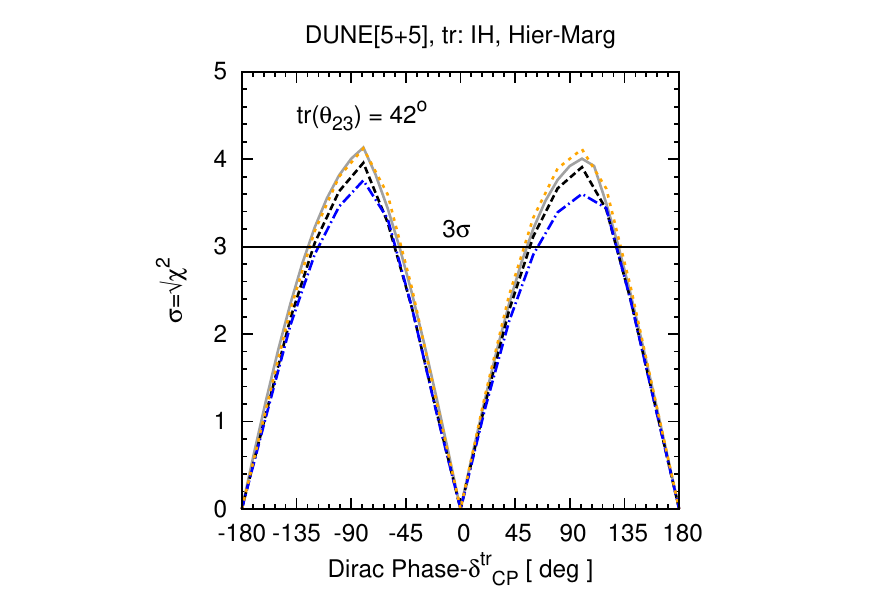}
\hspace{-3cm}
\includegraphics[height=6.5cm,width=10cm]{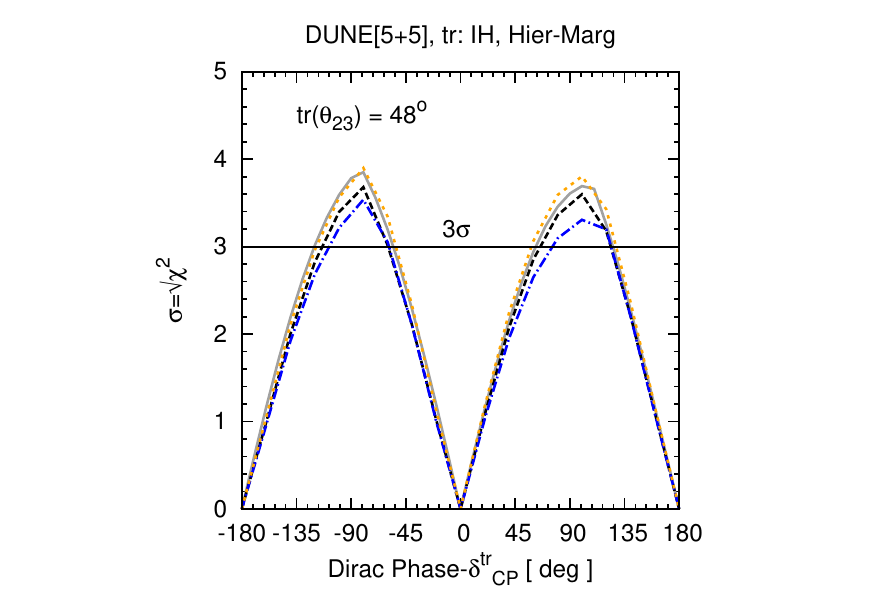}
 \end{tabular}
 \end{center}
\vspace{-4ex}        
\caption{\footnotesize CP-violation discovery $ \chi^{2} $ for DUNE in presence of NSI. Here top (bottom) row is for NH (IH) whereas in the left (right) panel, we show results for LO (HO).}
\label{fig:cpv_dis}
\end{figure}

In Fig. \ref{fig:cpv_dis}, we discuss CP-violation (CPV) discovery $\chi^{2} $ for DUNE considering both SI and NSI. The left (right) panel describes our result for LO (HO) whereas top (bottom) row discusses our results for true hierarchy as NH (IH).
The true values that we considered corresponding to 
$ \epsilon_{ee} = 0.095 $, $ \epsilon_{e \tau} = 0.0117 $ are based on our model as given in Tables~(\ref{nsiee}, and \ref{tab:nsiparam});
whereas for the new CP-phase due to NSI we take two cases, namely CP conserving ($ 0^\circ $) and CP violating ($ -90^\circ $) values. We describe CPV discovery $ \chi^{2} $ as,
\begin{equation}
\chi^{2}_{\rm CPV} = {\rm min}\sum_i \dfrac{ \left[  N_i(\delta^{\rm tr}_{CP}, \epsilon^{\rm tr}, \phi^{\rm tr})  - N_i(\delta^{\rm te}_{CP} (0^\circ, \pm 180^\circ) , \epsilon^{\rm te}, \phi^{\rm tr}) \right] ^{2}}{ \sigma [N_i(\delta^{\rm tr}_{CP}, \epsilon^{\rm tr}, \phi^{\rm tr})]^{2}} \; ,
\end{equation}
where $ N_i $ represents the number of events for the $i^{th}$ oscillation parameters.  Also $ \epsilon, \phi $ are defined as Eq.~\eqref{eq:mass_hier}.
We draw a $ 3\sigma  $ line as a benchmark for the discussion of our results. Comparing all curves, we notice that DUNE achieves maximum CPV discovery sensitivity for the SI compared to NSI parameters. Whereas, considering diagonal NSI parameter, we find that it gives almost equal CP sensitivity as SI at 3$\sigma$ C.L. for both the hierarchies. 
Further, when we add off-diagonal NSI parameters with the diagonal NSI parameter, as shown by blue dotted-dash and yellow dotted curves, we observe that CPV sensitivity reasonably decreases for NH. Whereas for IH, we find that  CPV sensitivity gets enhanced for CP violating value compare to CP conserving value.
Thus, an extra CP phase complicates the measurement of Dirac-CP phase ($ \delta_{CP} $) and hence effects the measurement of overall CP sensitivity even in the case of constraint parameter space of NSIs. 
%
 
\section{Summary and Conclusion}
\label{sec:concl}
Neutrino oscillation experiments opened a new vista to probe the fundamental properties of neutrinos. New physics models, incorporating neutrino masses, are testable in these experiments via the oscillation data. Many of these BSM scenarios give rise to NSI that can be tested in the oscillation experiments. But such models are constrained because of lepton flavor violation issues.   
The traditional $\nu$2HDM is one of these popular models which tries to explain neutrino masses by extending SM with two scalar doublets and right-handed neutrinos. However, this model produces negligible amount of NSI due to almost non-existent interaction of the SM charged leptons and quarks with the neutrinos, determined by the tiny mixing ($\approx v_1/v_2$) of the two scalar doublets. In this study, we propose a modified $\nu$2HDM which is an improvement over the usual $\nu$2HDM. We can have sizable NSI parameter while maintaining LFV constraints. We achieve this by assigning a charge to $e_R$ under a global $U(1)$ symmetry. This can lead to an observationally significant NSI along with the presence of a tiny Dirac neutrino mass, keeping the original motivation of $\nu$2HDM intact. This modification of $\nu$2HDM reintroduces hierarchy in the Yukawas in the $e- \nu$ sector, but simultaneously eases the hierarchy in the $t - \mu$ side, compared to those in the the traditional $\nu$2HDM. Softly broken global $U(1)$ $\nu$2HDM allows the presence of heavy neutral BSM scalar, that helps to address the stringent bound on electrophilic Yukawa of an ultra-light neutral scalar as well as the tight constraints from the oblique parameter ($S, T, U$) measurements. Combined effects of the LEP constraints like $e^+ e^- \to l^+ l^-$, mono-photon search along, with the bound on $H^{\pm}$ mass, reduces a significant amount of allowed parameter space of the modified $\nu$2HDM case, therefore putting stringent bounds on the NSI parameters. Presence of lepton flavor violating (LFV) decays like $\tau  \to 3 e$, $\mu \to 3 e $ put stringent upper bound on the Yukawa couplings $y_1, y_2$ and that results in any NSI parameter involving $y_1$ negligible apart from a significant reduction of the upper bound of other NSI parameters. 
Depending on these constraints, this model predicts the range of permissible NSI parameter $\epsilon_{ee}$. We also find that the only possible off-diagonal NSI parameter in this model is $\epsilon_{e \tau}$ whereas remaining NSIs become insignificant under model constraint. Thus, the effects of NSI parameters involving $y_1$ such as $\epsilon_{e \mu}, \epsilon_{\mu \mu} $ etc. is not studied here. Later, we study the impact of these NSIs considering LBL experiment like, DUNE. At the probability level, considering model-dependent NSIs, we observe no wrong hierarchy degeneracy even in the presence of off-diagonal NSI parameter. Furthermore, at the $ \chi^{2} $ level, we find that DUNE shows around 5$ \sigma $ hierarchy sensitivity when one adopt NH as well as IH both as a true hierarchy considering one at a time. These results remain valid irrespective of the value of true Dirac CP phase, $ \delta_{CP} $.  From our study of CP discovery, we notice that CP violation in leptonic sector gets affected even in the presence of model-dependent diagonal NSI parameter. Further, we observe that extra CP-phase, due to off-diagonal NSI parameter, complicates the measurement of Dirac-CP phase and hence affects the measurement of overall CP sensitivity.

\paragraph*{Acknowledgements\,:}
UKD acknowledges the support from Department of Science and Technology, Government of India under the fellowship reference number PDF/2016/001087 (SERB National Post-Doctoral Fellowship). UKD thanks Dr. Tirtha Sankar Ray for useful discussion. The research work of NN was supported in part by the National Natural Science Foundation of China under grant No.11775231. The authors would like to thank Prof. Srubabati Goswami for her insightful comments and careful reading of the manuscript.

\bibliographystyle{JHEP}
\bibliography{ref.bib}

\end{document}